\documentclass[%
reprint,
amsmath,amssymb,
aps,
pra,
]{revtex4-2}

\usepackage{graphicx}
\usepackage{dcolumn}
\usepackage{bm}
\usepackage{braket}
\usepackage{amsmath}
\usepackage{amsfonts}
\usepackage{multirow}
\usepackage{booktabs}
\usepackage{multirow}
\usepackage{tabularx}
\usepackage{array}

\begin{document}
	
	\preprint{APS/123-QED}
	
	\title{Quantum Computing for Electronic Circular Dichroism Spectrum Prediction of Chiral Molecules}

	\author{Amandeep Singh Bhatia}
	\email{amandeepbhatia.singh@gmail.com}
    \affiliation{ Department of Electrical and Computer Engineering, and Department  of Chemistry,  North Carolina State University, NC 27606, USA} 
	
	\author{Sabre Kais}%
	\email{Corresponding author: skais@ncsu.edu}
	\affiliation{ Department of Electrical and Computer Engineering, and Department  of Chemistry,  North Carolina State University, NC 27606, USA} 
	\begin{abstract}
		      Electronic circular dichroism (ECD) spectroscopy captures the chiroptical response of molecules, enabling absolute configuration assignment that is vital for enantioselective synthesis and drug design. The practical use of ECD spectra in predictive modeling remains restricted, as existing approaches offer limited confidence for chiral discrimination. By contrast, theoretical ECD calculations demand substantial computational effort rooted in electronic structure theory, which constrains their scalability to larger chemically diverse molecules. These limitations underscore the need for computational approaches that retain first-principles physical rigor while enabling efficient and scalable prediction. Motivated by recent advances in quantum algorithms for chemistry, we introduce a variational quantum framework combined with the quantum equation-of-motion formalism to compute molecular properties and predict ECD spectra, implemented within a multi-GPU/QPU accelerated hybrid quantum–classical workflow. We demonstrate its efficient  applicability on 12 clinically relevant chiral drug molecules accessing expanded active spaces. The proposed framework is assessed by comparison with established classical wavefunction-based methods, employing Coupled Cluster Singles and Doubles (CCSD) for ground-state energy benchmarks and Complete Active Space Configuration Interaction (CASCI)  as the reference method for excited-state energies and chiroptical properties within the same active orbital space. Notably, the quantum computed ECD spectra, obtained from chemically relevant active spaces mapped onto quantum circuits of approximately 20–24 qubits, exhibit near-quantitative agreement with classical reference calculations, accurately reproducing spectral line shapes, Cotton-effect signs, and relative peak intensities. These results establish quantum algorithms as a practical and scalable route to first-principles prediction of chiroptical spectra, opening the door to quantum-enabled molecular spectroscopy for chemically and pharmaceutically relevant systems.
              \vspace{1cm}

 
	\end{abstract}

	\maketitle
	\section{Introduction}

\begin{figure*}[!ht]
	\centering
\includegraphics[scale=0.65]{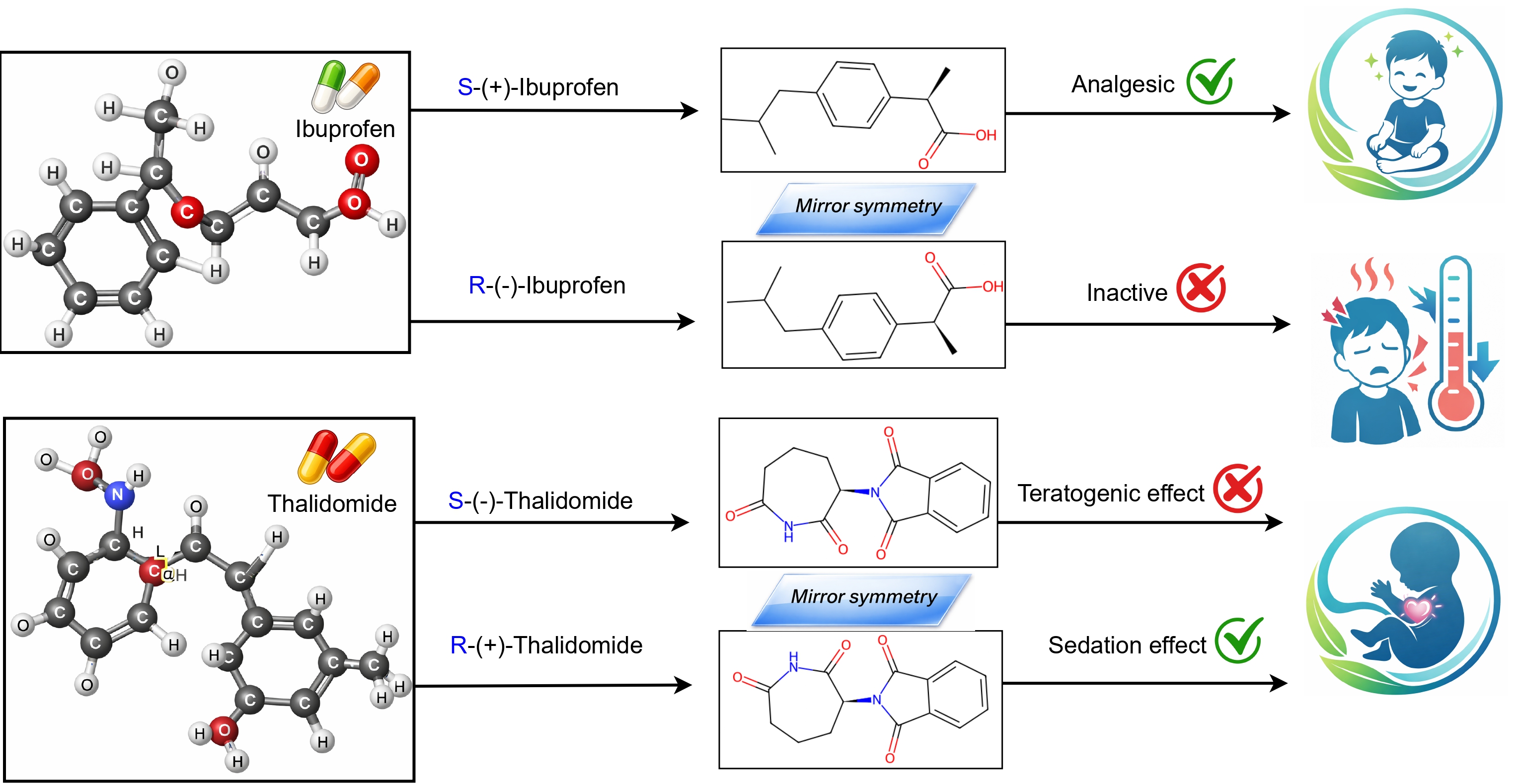}
\caption{\textbf{Enantioselectivity determines drug efficacy and safety.} Ibuprofen is a commonly prescribed nonsteroidal
anti-inflammatory drug used to treat pain, inflammation, and fever in both adults and pediatric patients, with therapeutic
activity arising from the (S)-enantiomer. Thalidomide displays opposite enantioselective effects, as its (R)-enantiomer exhibits
sedative effects in patients, whereas exposure to the (S)-enantiomer during pregnancy results in teratogenic effects affecting
fetal development.}
\label{medeffect}
\end{figure*}

    A chiral molecule exists as two mirror-related structures that share composition but differ in three-dimensional arrangement \cite{1}. In drug discovery, such structural differences can directly influence pharmacological behavior \cite{2}. In medicinal chemistry, such stereochemical differences can significantly affect therapeutic efficacy and safety \cite{3}. Ibuprofen, a widely used nonsteroidal anti-inflammatory drug, illustrates this principle, as its enantiomers display unequal biological activity \cite{4}. Specifically, the S-enantiomer is responsible for the desired anti-inflammatory effect, while the R-enantiomer shows substantially lower activity, emphasizing the importance of accurate stereochemical assignment in healthcare. Another clinically relevant example of chirality in medicine is Albuterol, a commonly prescribed bronchodilator used in the management of pediatric asthma \cite{5}. Its R-enantiomer is primarily responsible for therapeutic bronchodilation, whereas the S-enantiomer exhibits reduced efficacy and has been associated with adverse inflammatory responses, reinforcing the importance of enantiomer-specific characterization in treatments for children. These examples highlight the critical role of accurate absolute configuration determination in chiral research and its direct implications for safe and effective therapeutic design \cite{6}. Fig~\ref{medeffect} illustrates the central role of molecular chirality in determining drug efficacy and safety. The examples highlight how distinct enantiomers of the same compound can exhibit markedly different, and sometimes opposing, biological activities, underscoring the importance of reliable chiral discrimination in pharmaceutical research. These well-established cases motivate the need for accurate methods to assign absolute configuration and predict chiroptical response, providing the chemical and biomedical context for the ECD-based approach developed in this work.

Reliable chiral assignment is fundamental to applications ranging from asymmetric catalysis to functional materials and pharmaceuticals. The experimental chiroptical measurements alone are insufficient to uniquely determine absolute configuration, reliable theoretical modeling is essential. Electronic circular dichroism (ECD) is extensively used because it probes electronic transitions that are highly sensitive to molecular chirality and can be directly compared with quantum-chemical predictions \cite{7, 8}. Other approaches, including vibrational circular dichroism (VCD), Raman optical activity (ROA), X-ray crystallography, and NMR-based chiral analysis, can also provide stereochemical information but are often limited by sample requirements, structural constraints, or reduced applicability to complex molecular systems \cite{9, 10}. 




\begin{figure*}[!ht]
	\centering
\includegraphics[scale=0.65]{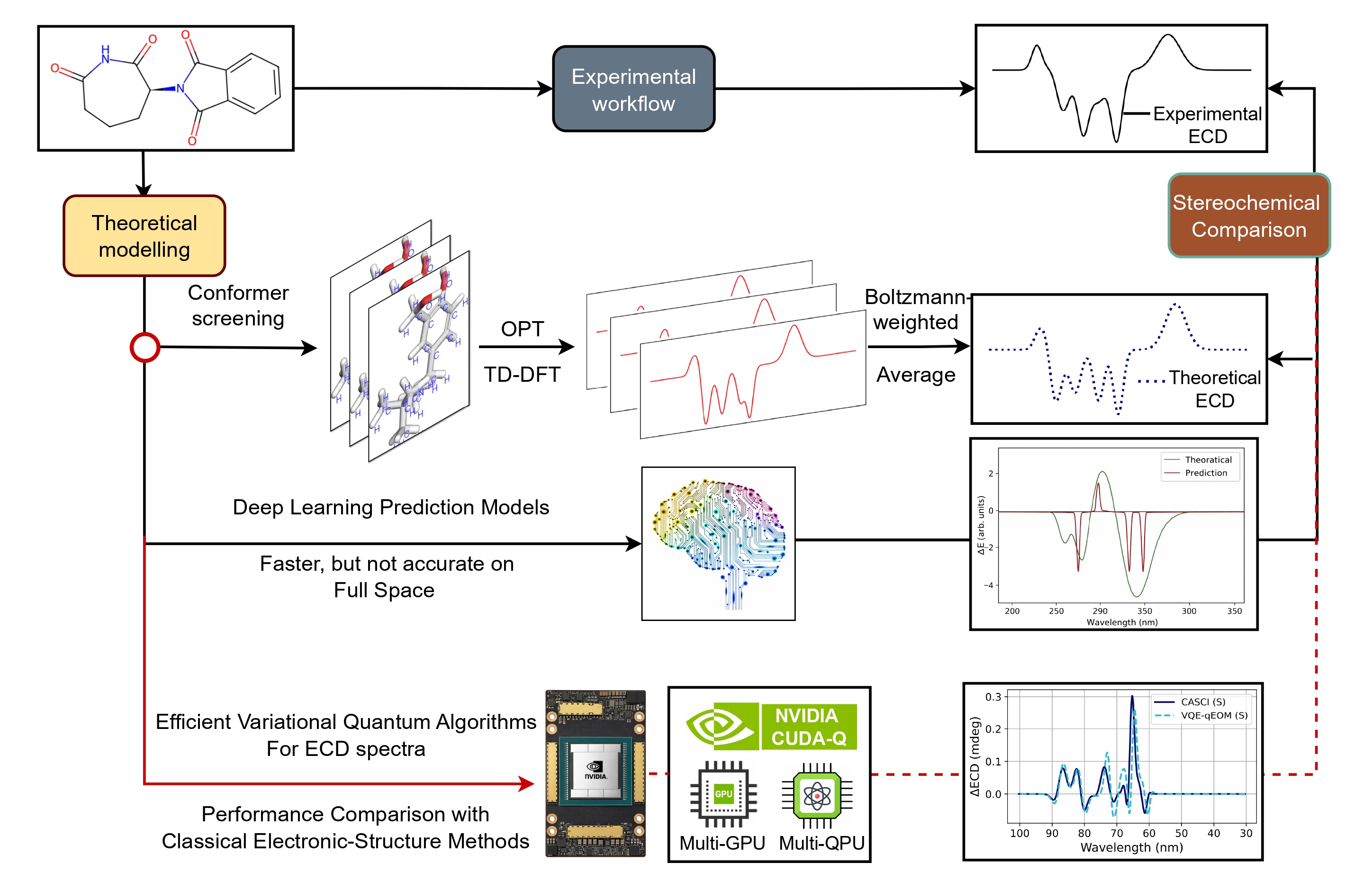}
\caption{\textbf{The Schematic representation of the ECD prediction and chiral assignment.} Electronic Circular Dichroism (ECD) comparison is widely used for absolute configuration assignment, but theoretical ECD simulations are computationally demanding due to conformational screening, geometry optimization, excited-state calculations, and Boltzmann-weighted averaging. Deep-learning–based approaches enable accelerated ECD prediction but often lack quantitative accuracy and transferability; this motivates the variational quantum algorithm presented here, which computes ECD spectra using multi-GPU- and QPU-accelerated quantum–classical workflows within an active-space framework and is validated against CASCI.}
\label{mainfig}
\end{figure*}

The calculation of ECD spectra for chiral molecules requires a layered theoretical procedure involving both structural and electronic analysis \cite{11}. Following molecular construction, extensive conformational sampling is performed to identify low-energy geometries. These structures are then refined using density functional theory (DFT) and their ECD signatures are calculated through time-dependent electronic structure techniques \cite{12}. A population-weighted average of the conformer-specific spectra yields the final theoretical ECD response. The need for specialized expertise and the substantial computational resources associated with this process make it a significant bottleneck in routine stereochemical assignment. Consequently, the conventional approach requires accurate simulations of ECD spectra, leading to high computational cost and limited scalability \cite{13, 14}. These challenges become increasingly pronounced for large molecular systems and conformationally flexible compounds, where extensive sampling and excited-state calculations are required. Consequently, the development of efficient and scalable theoretical approaches for chiroptical property prediction remains a pressing objective in chiral research.

In recent years, machine-learning–based statistical methods have been incorporated into chemical research workflows to accelerate chiroptical property prediction, including models that combine graph neural networks with transformer architectures. While such approaches offer significant computational efficiency, their performance remains highly sensitive to data quality, spectral diversity, and model generalization \cite{14}. Moreover, the wide variability in ECD spectral line shapes across molecular conformers and chemical classes complicates the extraction of robust latent representations, limiting the reliability of deep-learning models for absolute configuration assignment. The exact treatment of electronic structure and molecular dynamics is an NP-hard problem, implying that classical algorithms scale exponentially with system size \cite{15}. As a consequence, classical computational approaches are expected to become impractical for large or strongly correlated molecular systems due to prohibitive resource requirements.

In parallel with classical advances, quantum computing has seen substantial progress, especially in the formulation of algorithms targeting quantum chemistry and materials science problems on future fault-tolerant devices \cite{16, 17, 18, 19}. Notably, a range of quantum algorithms, including quantum phase estimation (QPE) \cite{20, 21} , variational quantum eigensolvers (VQE) \cite{22, 23, 24}, quantum machine leraning based on Restricted Boltzman Machine (RBM) \cite{RBM-1,RBM-2,RBM-3}, equation-of-motion–based extensions, quantum imaginary time evolution (QITE) \cite{25, 26}, and quantum algorithms for linear systems have been developed for electronic structure calculations \cite{27, 28}, molecular response property evaluation, and the simulation of strongly correlated systems, with the potential to offer runtime advantages over classical approaches in specific regimes. With recent progress in quantum hardware development, increasing attention has been directed toward leveraging near-term quantum devices for the evaluation of molecular properties.

In particular, researchers have explored hybrid quantum–classical strategies that balance hardware limitations with algorithmic robustness, enabling meaningful chemical insights on noisy intermediate-scale quantum (NISQ) platforms \cite{29}. These efforts aim to bridge the gap between current hardware capabilities and the long-term promise of fault-tolerant quantum computation. The authors \cite{101} use hybrid quantum--classical algorithms, including VQE--qEOM and
stochastic quantum dynamics (SQD), to accurately predict excited-state spectra of
common battery electrolyte salts within NISQ-scale models. Their results capture
systematic excitation-energy trends across both anions and cations, providing
quantitative guidance for electrolyte design. The authors \cite{102} propose an orbital-optimized VQE–qEOM framework for computing
excitation energies and spectroscopic properties on near-term quantum devices.
Benchmark simulations demonstrate agreement with classical CASSCF for small
molecular systems. Recently, the authors \cite{103} introduce a particle-number-conserving multi-reference unitary coupled-cluster (MR-UCC) algorithm that achieves accurate ground-state energies with minimal quantum resources using a single circuit across all bond lengths. Grimsley et al. \cite{104} introduce a multistate adaptive VQE approach that accurately computes molecular excited states and transition properties, outperforming single-state ADAPT-VQE and q-sc-EOM in describing state crossings and avoided crossings in strongly correlated systems.

Fig.~\ref{mainfig} outlines the computational strategy used for ECD prediction and chiral assignment. The schematic contrasts the complexity of conventional theoretical ECD workflows with a hybrid quantum–classical formulation in which excited-state chiroptical properties are evaluated within an active-space framework. Classical pre- and post-processing are accelerated using multi-GPU resources, while the variational quantum eigensolver and quantum equation-of-motion components are executed on quantum processing units. This integrated approach enables efficient and quantitatively reliable ECD simulations that are subsequently benchmarked against classical CASCI references.

Extensive benchmarking efforts in classical quantum chemistry have established both high-accuracy and cost-effective methods for computing excitation energies of small to medium-sized organic molecules, ranging from coupled-cluster approaches to large curated benchmark databases. The multiconfigurational problems are commonly addressed using active-space methods such as CASCI \cite{30}, CASSCF \cite{31}, and DMRG \cite{32}, whereas their quantum analogues, including orbital-optimized variational approaches, offer a potential pathway to treating larger active spaces by offloading the exponential cost of wavefunction optimization to quantum hardware \cite{33, 34, 35}. Active-space formulations are naturally aligned with quantum computing, as they isolate the orbitals responsible for strong correlation and chemical activity \cite{39}. This restriction substantially reduces qubit requirements and circuit complexity, enabling feasible quantum simulations while allowing direct, controlled comparisons with classical multiconfigurational methods.

These classsical approaches can be recast within the unitary coupled-cluster (UCC) formalism \cite{36}, which enforces unitarity and provides a variational description of correlated electronic structure, making it well suited for multiconfigurational systems encountered in chemistry and biology. Practical realizations include variants such as UCCSD, UCCD \cite{37}, PUCCSD \cite{38} and UVCCSD \cite{24}, which differ in the types of excitation operators retained. Despite their favorable theoretical properties, many UCC based ansätze remain associated with substantial circuit depth, motivating the development and selection of more compact formulations. The unitary coupled-cluster (UCC) wavefunction is defined as
\begin{equation}
\label{eq:ucc}
\ket{\Psi_{\mathrm{UCC}}}
=
e^{\hat{T} - \hat{T}^{\dagger}} \ket{\Phi_0},
\end{equation}
where $\ket{\Phi_0}$ denotes a reference state, typically the Hartree-Fock determinant, and
$\hat{T}$ is the cluster excitation operator expressed as a sum of particle-hole excitations,
\begin{equation}
\hat{T} = \hat{T}_1 + \hat{T}_2 + \hat{T}_3 + \cdots .
\end{equation}
Here, the single and double-excitation operators are given by
\begin{equation}
\hat{T}_1 = \sum_{i,a} t_i^{a}\, a_a^{\dagger} a_i,
\qquad
\hat{T}_2 = \sum_{i<j,\,a<b} t_{ij}^{ab}\, a_a^{\dagger} a_b^{\dagger} a_j a_i,
\end{equation}
where indices $i,j$ label occupied orbitals, $a,b$ label virtual orbitals,
$a^\dagger$ and $a$ are fermionic creation and annihilation operators, and
$t_i^a$ and $t_{ij}^{ab}$ are variational cluster amplitudes.
The anti-Hermitian generator $\hat{T}-\hat{T}^\dagger$ ensures unitarity of the ansatz,
enabling a variational optimization on quantum computing platforms.

\subsection{Current challenges}
In practice, Time-Dependent DFT (TD-DFT) evaluation of ECD spectra is the principal rate determining component, underscoring the challenge and importance of improving computational efficiency \cite{11}. Beyond data-driven limitations, conventional first-principles approaches to ECD prediction face substantial computational challenges. Accurate simulation of ECD spectra typically requires high-level excited-state electronic structure methods combined with extensive conformational sampling, leading to steep computational scaling with molecular size. These demands become prohibitive for large, flexible, or biologically relevant molecules, restricting routine application in pharmaceutical settings.

Additional challenges arise from the sensitivity of ECD spectra to electronic correlation effects,  and methodological approximations, which can result in inconsistent predictions across different theoretical frameworks \cite{11, 40}. Moreover, the lack of systematic error control and interpretability in both classical and machine-learning approaches complicates the assessment of confidence in absolute configuration assignments \cite{14}. Together, these issues underscore the need for scalable, physically grounded methodologies capable of delivering reliable and interpretable chiroptical predictions.

\subsection{Motivation}
The limitations inherent to data-driven approaches underscore the need for physically grounded methodologies capable of accurately and efficiently predicting chiroptical properties. The traditional workflow for absolute configuration assignment relies heavily on high-level theoretical simulations of ECD spectra, which are computationally intensive and time-consuming. In practice, this process involves multiple labor-intensive steps, including enantiomeric separation, acquisition of experimental circular dichroism data, and quantum-chemical computation of reference ECD spectra to reach a definitive stereochemical assignment.

Quantum computing presents a compelling alternative by enabling first-principles simulations of molecular electronic structure without reliance on large empirical training datasets. In this context, quantum algorithms offer a principled framework for modeling electronic response properties through an explicit treatment of electronic correlation effects. It has the potential to alleviate the computational demands associated with conventional chiroptical simulations while preserving first-principles accuracy. Moreover, quantum computing based prediction of ECD spectra provides an interpretable and transferable pathway for absolute configuration determination, particularly for chiral molecules of biomedical relevance. Accordingly, this work is motivated by the development of scalable, physics based quantum methodologies that complement and extend existing classical machine learning approaches to chiroptical spectroscopy.

\subsection{Main contributions}
To summarize, this work makes the following contributions:

\begin{itemize}
    \item We introduce a variational quantum framework for predicting electronic circular dichroism  spectra of pharmaceutically relevant chiral molecules, enabling absolute configuration assignment without reliance on large data-driven or empirical models.
    \item We develop an end-to-end quantum--classical workflow that links electronic structure modeling to ECD spectra through a fully quantum-consistent treatment of ground- and excited-state properties.
    \item We provide a systematic validation of the proposed approach against established classical electronic-structure methods within well-defined active spaces, demonstrating quantitative agreement for both single and multi-chiral-center drug molecules of direct relevance to pharmaceutical and healthcare applications.
\end{itemize}

The remainder of the article is organized as follows. Section 2 introduces the theoretical framework and computational methodology underlying the quantum-based prediction of chiroptical properties. Section 3 describes the experimental and computational setup, including the selection of healthcare-relevant chiral drug molecules and implementation details. Section 4 presents the results and provides a detailed discussion of the predicted electronic circular dichroism spectra and rotatory strengths. Finally, Section 5 concludes the paper with a summary of key findings and outlines potential directions for future research.

\section{Theory}

\subsection{Active-space approximation}

The active-space approximation is constructed based on the one-particle reduced density matrix (1-RDM) obtained from a correlated electronic structure calculation. The spin-summed 1-RDM is defined as
\begin{equation}
\gamma_{pq}
=
\langle \Psi \,|\, \hat{a}_p^{\dagger} \hat{a}_q \,|\, \Psi \rangle,
\end{equation}
where $\hat{a}_p^{\dagger}$ and $\hat{a}_q$ are fermionic creation and annihilation operators, and $|\Psi\rangle$ denotes an approximate correlated many-electron wavefunction.

Diagonalization of the 1-RDM yields the natural orbitals $\{\phi_i\}$ and their corresponding natural orbital occupation numbers $\{n_i\}$,
\begin{equation}
\sum_q \gamma_{pq}\,\phi_i(q) = n_i\,\phi_i(p),
\end{equation}
with occupation numbers satisfying
\begin{equation}
0 \le n_i \le 2 .
\end{equation}

Orbitals with $n_i \approx 2$ correspond to doubly occupied core orbitals, while orbitals with $n_i \approx 0$ correspond to unoccupied virtual orbitals. Orbitals exhibiting fractional occupation numbers ($0 < n_i < 2$) indicate significant static or near-degeneracy correlation and are essential for a multiconfigurational description.

The active space $\mathcal{A}$ is defined by retaining orbitals whose occupation numbers deviate from the limiting values,
\begin{equation}
\mathcal{A}
=
\left\{ \phi_i \; \big| \; \epsilon < n_i < 2 - \epsilon \right\},
\end{equation}
where $\epsilon$ is a threshold ($\sim0.02$) controlling the balance between physical accuracy and computational feasibility. The remaining orbitals are treated as frozen core or external virtual orbitals.

\subsection{Hamiltonian formulation and qubit encoding}
Following the construction of the active orbital subspace, the electronic structure problem is reformulated within this reduced Hilbert space. The ground-state energy is obtained variationally by minimizing the expectation value of the electronic Hamiltonian with respect to a parametrized trial wavefunction,
\begin{equation}
E(\boldsymbol{\theta})
=
\langle \Psi(\boldsymbol{\theta}) \,|\, \hat{H}_{\mathrm{act}} \,|\, \Psi(\boldsymbol{\theta}) \rangle ,
\end{equation}
which is bounded from below by the exact ground-state energy.

Within the Born-Oppenheimer approximation, the active-space electronic Hamiltonian is expressed in second-quantized form as
\begin{equation}
\hat{H}_{\mathrm{act}}
=
\sum_{pq} h_{pq}\, \hat{a}_p^{\dagger} \hat{a}_q
+
\frac{1}{2}
\sum_{pqrs} h_{pqrs}\, \hat{a}_p^{\dagger} \hat{a}_q^{\dagger} \hat{a}_r \hat{a}_s ,
\end{equation}
where indices $p,q,r,s$ run over active molecular orbitals, $\hat{a}_p^{\dagger}$ and $\hat{a}_p$ denote fermionic creation and annihilation operators, and $h_{pq}$ and $h_{pqrs}$ are one and two-electron integrals evaluated in the active orbital basis. These integrals are obtained from classical electronic structure calculations.

To enable quantum simulation, the fermionic Hamiltonian is mapped onto a qubit representation using a fermion-to-qubit transformation, resulting in
\begin{equation}
\hat{H}_{\mathrm{act}}
=
\sum_k c_k \, \hat{P}_k
=
\sum_k c_k \bigotimes_{j=1}^{N_q} \sigma^{(k)}_j ,
\end{equation}
where, $c_k \in \mathbb{R}$ are real-valued coefficients determined by the one and two-electron integrals of the active-space Hamiltonian, $\hat{P}_k$ denote Pauli strings, and $N_q$ is the total number of qubits, equal to the number of active spin orbitals. The operator $\sigma^{(k)}_j \in \{I,X,Y,Z\}$ represents the Pauli operator acting on qubit $j$ in the $k$-th term, while the tensor product $\bigotimes_{j=1}^{N_q}$ indicates that each Pauli string acts non-trivially on a subset of qubits and as the identity on the remaining qubits. This qubit Hamiltonian provides a complete qubit-level representation of the active-space electronic Hamiltonian and serves as the input for subsequent variational quantum simulations.

\begin{figure*}[!ht]
	\centering
\includegraphics[scale=0.22]{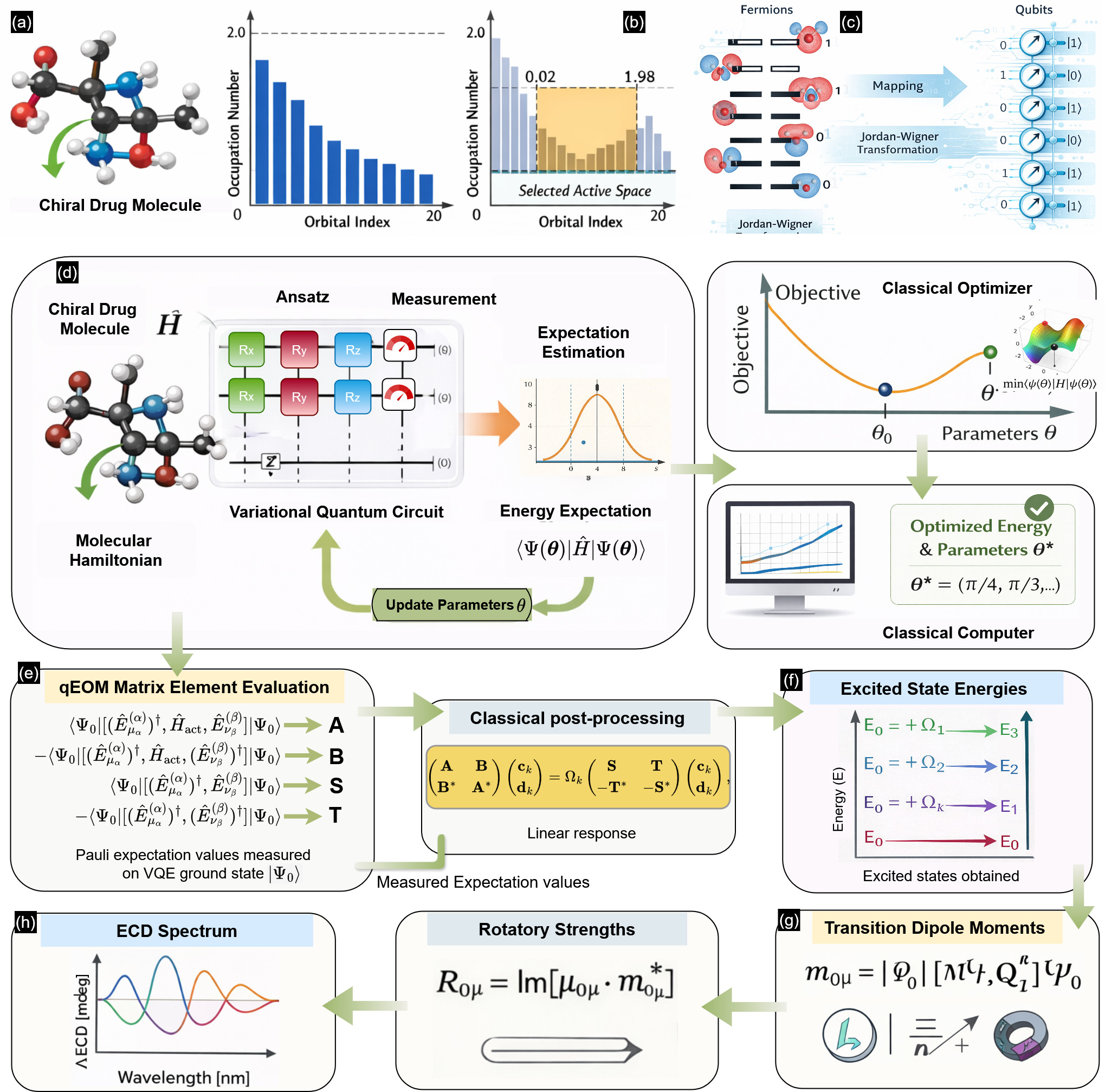}
\caption{\textbf{Quantum-classical workflow for computing electronic circular dichroism (ECD) spectra of chiral drug molecules.} (a) Molecular structure of a representative chiral pharmaceutical compound.
(b) Selection of the active orbital subspace based on natural occupation numbers, retaining orbitals with occupations in the range
$0.02 \le n_i \le 1.98$
to capture essential static and dynamic correlation effects.
(c) Construction of the active-space electronic Hamiltonian and its mapping to a qubit representation using fermionic second-quantized operators followed by the Jordan--Wigner transformation.
(d) Ground-state energy optimization using the VQE, where a parametrized quantum circuit is iteratively optimized by a classical optimizer until convergence.
(e-f) Evaluation of excited-state energies via the qEOM formalism built upon the converged VQE reference state.
(g) Computation of electric and magnetic transition dipole moments and rotatory strengths from qEOM excited states, (h) leading to the final ECD spectrum.
}
\label{flow}
\end{figure*}

\subsection{Variational Quantum EigenSolver}
Variational Quantum EigenSolver (VQE) is a hybrid quantum–classical algorithm that leverages the variational principle to estimate the lowest eigenvalue of a many-body Hamiltonian using a parameterized family of quantum states. Given a qubit-encoded Hamiltonian $\hat{H}_{\mathrm{act}}$, the algorithm approximates the ground state by preparing a parametrized quantum state on a quantum processor and iteratively optimizing its parameters using a classical optimizer.

In VQE, a trial wavefunction is prepared by applying a parametrized quantum circuit (ansatz) $\hat{U}(\boldsymbol{\theta})$ to an initial reference state $\lvert \Phi_0 \rangle$, typically chosen as a computational basis state or a mean-field state,
\begin{equation}
\lvert \Psi(\boldsymbol{\theta}) \rangle
=
\hat{U}(\boldsymbol{\theta}) \lvert \Phi_0 \rangle ,
\end{equation}
where $\boldsymbol{\theta} = (\theta_1, \theta_2, \ldots)$ denotes a set of variational parameters controlling the single and multi-qubit gates within the circuit. The form of the ansatz may be problem-inspired, such as unitary coupled-cluster constructions, or hardware-efficient, consisting of alternating layers of parameterized single-qubit rotations and entangling gates.

The energy associated with the trial state is evaluated as the expectation value of the qubit Hamiltonian,
\begin{equation}
E(\boldsymbol{\theta})
=
\langle \Psi(\boldsymbol{\theta}) \lvert \hat{H}_{\mathrm{act}} \rvert \Psi(\boldsymbol{\theta}) \rangle
=
\sum_k c_k
\langle \Psi(\boldsymbol{\theta}) \lvert \hat{P}_k \rvert \Psi(\boldsymbol{\theta}) \rangle .
\end{equation}
Since each Pauli string $\hat{P}_k$ is a tensor product of single-qubit Pauli operators, its expectation value can be estimated efficiently on a quantum device by repeated projective measurements in appropriate measurement bases. The total energy is obtained by summing the measured expectation values weighted by their corresponding coefficients $c_k$.

The variational parameters are optimized on a classical computer by minimizing the measured energy,
\begin{equation}
E_0
=
\min_{\boldsymbol{\theta}} \, E(\boldsymbol{\theta}),
\end{equation}
which, by the variational principle, provides an upper bound to the exact ground-state energy of the Hamiltonian. This optimization is carried out iteratively for a given parameter set $\boldsymbol{\theta}$, the quantum processor prepares $\lvert \Psi(\boldsymbol{\theta}) \rangle$ and estimates the energy, while a classical optimizer updates $\boldsymbol{\theta}$ based on the energy values (and optionally gradients) until convergence is achieved.

The VQE framework thus combines quantum resources, used to represent and measure the many-body wavefunction, with classical optimization techniques, enabling the treatment of correlated electronic structure problems within active spaces that are challenging for purely classical methods. This hybrid approach makes VQE particularly well suited for near-term quantum devices and forms the basis for quantum simulations of molecular systems within the active-space formalism.

\subsection{Quantum Equation-of-Motion formalism for excited states}

While the VQE algorithm provides access to the ground-state energy and wavefunction of the active-space Hamiltonian, the description of excited states requires an extension beyond ground-state variational optimization. The Quantum Equation-of-Motion (qEOM) formalism achieves this by formulating excited states as linear response modes around a correlated reference state prepared by VQE, closely analogous to the classical equation-of-motion and linear-response theories used in quantum chemistry.

Let $\lvert \Psi_0 \rangle$ denote the approximate ground state obtained from the VQE, with corresponding energy $E_0$. Within the qEOM framework, excited states are generated by the action of excitation operators on the reference state,
\begin{equation}
\lvert \Psi_\mu \rangle
=
\hat{O}_\mu^\dagger \lvert \Psi_0 \rangle ,
\end{equation}
where the operators $\hat{O}_\mu^\dagger$ are chosen from a truncated excitation manifold.

In this work, we restrict the excitation operator basis to single $(\alpha=1)$ and double $(\alpha=2)$ excitations,
\begin{align}
\hat{E}^{(1)}_{\mu_1} &= \hat{a}_m^\dagger \hat{a}_i , \\
\hat{E}^{(2)}_{\mu_2} &= \hat{a}_m^\dagger \hat{a}_n^\dagger \hat{a}_i \hat{a}_j ,
\end{align}
with corresponding de-excitation operators
\begin{align}
\left( \hat{E}^{(1)}_{\mu_1} \right)^\dagger &= \hat{a}_i^\dagger \hat{a}_m , \\
\left( \hat{E}^{(2)}_{\mu_2} \right)^\dagger &= \hat{a}_i^\dagger \hat{a}_j^\dagger \hat{a}_m \hat{a}_n .
\end{align}
Here, indices $i,j$ denote occupied orbitals and $m,n$ denote virtual orbitals within the chosen active space.

The excitation energies $\omega_\mu = E_\mu - E_0$ are obtained by enforcing the equation-of-motion condition
\begin{equation}
\left[ \hat{H}_{\mathrm{act}}, \hat{O}_\mu^\dagger \right]
\lvert \Psi_0 \rangle
=
\omega_\mu \, \hat{O}_\mu^\dagger
\lvert \Psi_0 \rangle ,
\end{equation}
which ensures that the excited states satisfy the Schrödinger equation to first order in the excitation operators.

By expanding the excitation operators in the restricted basis and projecting onto the corresponding de-excitation space, the qEOM condition yields a generalized secular equation of the form
\begin{equation}
\begin{pmatrix}
\mathbf{A} & \mathbf{B} \\
\mathbf{B}^\ast & \mathbf{A}^\ast
\end{pmatrix}
\begin{pmatrix}
\mathbf{c}_k \\
\mathbf{d}_k
\end{pmatrix}
=
\Omega_k
\begin{pmatrix}
\mathbf{S} & \mathbf{T} \\
-\mathbf{T}^\ast & -\mathbf{S}^\ast
\end{pmatrix}
\begin{pmatrix}
\mathbf{c}_k \\
\mathbf{d}_k
\end{pmatrix},
\end{equation}
where $\mathbf{c}_k$ and $\mathbf{d}_k$ denote the forward and backward excitation amplitudes associated with the $k$th excited state, respectively.

The matrix elements are evaluated as expectation values with respect to the VQE-optimized ground state $\lvert \Psi_0 \rangle$ and are defined as
\begin{align}
A_{\mu_\alpha \nu_\beta}
&=
\langle \Psi_0 \lvert
\big[
(\hat{E}^{(\alpha)}_{\mu_\alpha})^\dagger ,
\hat{H}_{\mathrm{act}},
\hat{E}^{(\beta)}_{\nu_\beta}
\big]
\rvert \Psi_0 \rangle , \\
B_{\mu_\alpha \nu_\beta}
&=
- \langle \Psi_0 \lvert
\big[
(\hat{E}^{(\alpha)}_{\mu_\alpha})^\dagger ,
\hat{H}_{\mathrm{act}},
(\hat{E}^{(\beta)}_{\nu_\beta})^\dagger
\big]
\rvert \Psi_0 \rangle , \\
S_{\mu_\alpha \nu_\beta}
&=
\langle \Psi_0 \lvert
\big[
(\hat{E}^{(\alpha)}_{\mu_\alpha})^\dagger ,
\hat{E}^{(\beta)}_{\nu_\beta}
\big]
\rvert \Psi_0 \rangle , \\
T_{\mu_\alpha \nu_\beta}
&=
- \langle \Psi_0 \lvert
\big[
(\hat{E}^{(\alpha)}_{\mu_\alpha})^\dagger ,
(\hat{E}^{(\beta)}_{\nu_\beta})^\dagger
\big]
\rvert \Psi_0 \rangle .
\end{align}

Solving this generalized eigenvalue problem yields the excitation energies $\Omega_k$ and the corresponding eigenvectors $(\mathbf{c}_k, \mathbf{d}_k)$, which define the expansion of the excited-state wavefunctions within the truncated singles--doubles excitation manifold.

In practical quantum implementations, all matrix elements are evaluated by measuring expectation values of Pauli operator strings on the VQE-prepared ground state $\lvert \Psi_0 \rangle$. This approach avoids the explicit preparation of excited states and enables access to low-lying excitation energies using the same quantum circuit optimized for the ground state.

Fig~\ref{flow} summarizes the quantum–classical workflow used to compute ECD spectra of chiral drug molecules. The approach combines active-space Hamiltonian construction with variational quantum algorithms for ground- and excited-state calculations, followed by the evaluation of chiroptical observables. By integrating classical preprocessing and post-processing with quantum simulations, the workflow enables a systematic and scalable route to ECD prediction for pharmaceutically relevant chiral systems. The qEOM approach therefore provides an efficient and systematically improvable framework for computing low-lying excited states within the active-space Hamiltonian. 

\begin{table*}[t]
\centering
\caption{Chiral drug molecules investigated in this study, with representative SMILES strings specifying the stereochemical configuration of each enantiomer.}
\label{chiral_list}
\renewcommand{\arraystretch}{1.2}
\begin{tabular}{clll}
\hline
\textbf{Chirality Type} & \textbf{Drug Name} & \textbf{SMILES} & \textbf{Therapeutic Use} \\
\hline
\multirow{5}{*}{Single}
& Dimercaprol & OC[C@@H](S)CS & Chelation (arsenic, mercury \\ & & & poisoning) \\
& Phenprocoumon & C[C@H](CC)c1ccccc1C2=C(O)c3ccccc3OC2=O & Anticoagulant  \\ &  & & (prevention of thrombosis) \\
& Ketoprofen & C[C@H](C(=O)O)c1ccc(cc1)C(=O)c2ccccc2 & NSAID (pain, \\ & & &  inflammation,  and arthritis) \\
& Captopril & C[C@H](CS)C(=O)N1CCC[C@H]1C(=O)O  & ACE inhibitor (hypertension \\
& & & and heart failure) \\
& Omeprazole & COc1ccc2nc(S(=O)[C@](C)(C)c3nccnc3)sc2c1 & Proton pump inhibitor \\
& & & (GERD and peptic ulcers) \\
& Thalidomide & O=C1CC[C@@H](N2C(=O)c3ccccc3C2=O)C(=O)N1  & Immunomodulatory agent \\ & & &(leprosy complications) \\

\hline
\multirow{5}{*}{Multiple}
& Threonine & C[C@H]([C@@H](C(=O)O)N)O & Essential amino acid (nutrition \\ & & & and metabolism) \\
& Ephedrine & CN[C@H](C)[C@@H](O)c1ccccc1 & Sympathomimetic agent \\ & & & (bronchodilation, decongestant) \\
& D-Ribose & OC1[C@@H](CO)[C@H](O)[C@H](O)O1 & Metabolic sugar \\ & & & (nucleotide and energy metabolism) \\
& Menthol & CC(C)[C@@H]1CC[C@@H](C)C[C@@H]1O & Analgesic and cooling agent \\ & & &(topical applications) \\
& Isosorbide & C1[C@H]([C@@H]2[C@H](O1)[C@H](CO2)O)O & Osmotic agent \\ & & & (glaucoma and intracranial pressure) \\
& Borneol & CC1(C)[C@@H]2CC[C@@]1(C)[C@@H](O)C2 & Terpenoid compound (topical  \\ & & & analgesic and penetration enhancer) \\

\hline
\end{tabular}
\end{table*}

\subsection{Chiroptical properties}

In this paper, the qEOM approach builds upon the correlated VQE ground state to capture both static and dynamic correlation effects, thereby providing access to excitation energies and spectroscopic observables within the capabilities of near-term quantum hardware.

Chiroptical observables, such as electronic circular dichroism (ECD), arise from the coupled response of electric and magnetic transition moments associated with electronic excitations. Within the present framework, these quantities are evaluated using excited states obtained from the qEOM formalism, enabling the computation of excitation energies and chiroptical response properties while consistently incorporating both static and dynamic electron correlation effects.

\begin{table*}[t]
\centering
\caption{Ground-state energies (in Hartree) of chiral drug molecules computed using CASCI, CCSD, and VQE--qEOM methods.}
\label{gse}
\renewcommand{\arraystretch}{1.25}

\begin{tabular}{lccc|lccc}
\hline
\multicolumn{4}{c|}{\textbf{Single chiral center}} 
& \multicolumn{4}{c}{\textbf{Multiple chiral centers}} \\
\hline
\textbf{Drug Name} & \textbf{CASCI} & \textbf{CCSD} & \textbf{VQE}
& \textbf{Drug Name} & \textbf{CASCI} & \textbf{CCSD} & \textbf{VQE} \\
\hline
Dimercaprol     & -977.04 & -977.04 & -977.02 & Threonine    & -430.11 & -430.10 & -430.08 \\
Phenprocoumon   & -942.67 & -942.64 & -942.65 & Ephedrine    & -510.39 & -510.28 & -509.81 \\
Ketoprofen       & -828.11 & -828.08 & -828.08 & Ribose       & -562.05 & -562.04 & -562.04 \\
Captopril       & -1013.82 & -1013.79 & -1013.81 & Menthol      & -459.60 & -459.59 & -459.59 \\
Omeprazole      & -1664.89 & -1664.88 & -1664.86 & Isosorbide   & -525.64 & -525.64 & -525.58 \\
Thalidomide        & -894.97 & -894.95 & -894.92 & Borneol  & -458.44 & -458.44 & -458.44 \\
\hline
\end{tabular}
\end{table*}

\begin{figure*}[!ht]
	\centering
\includegraphics[scale=0.6]{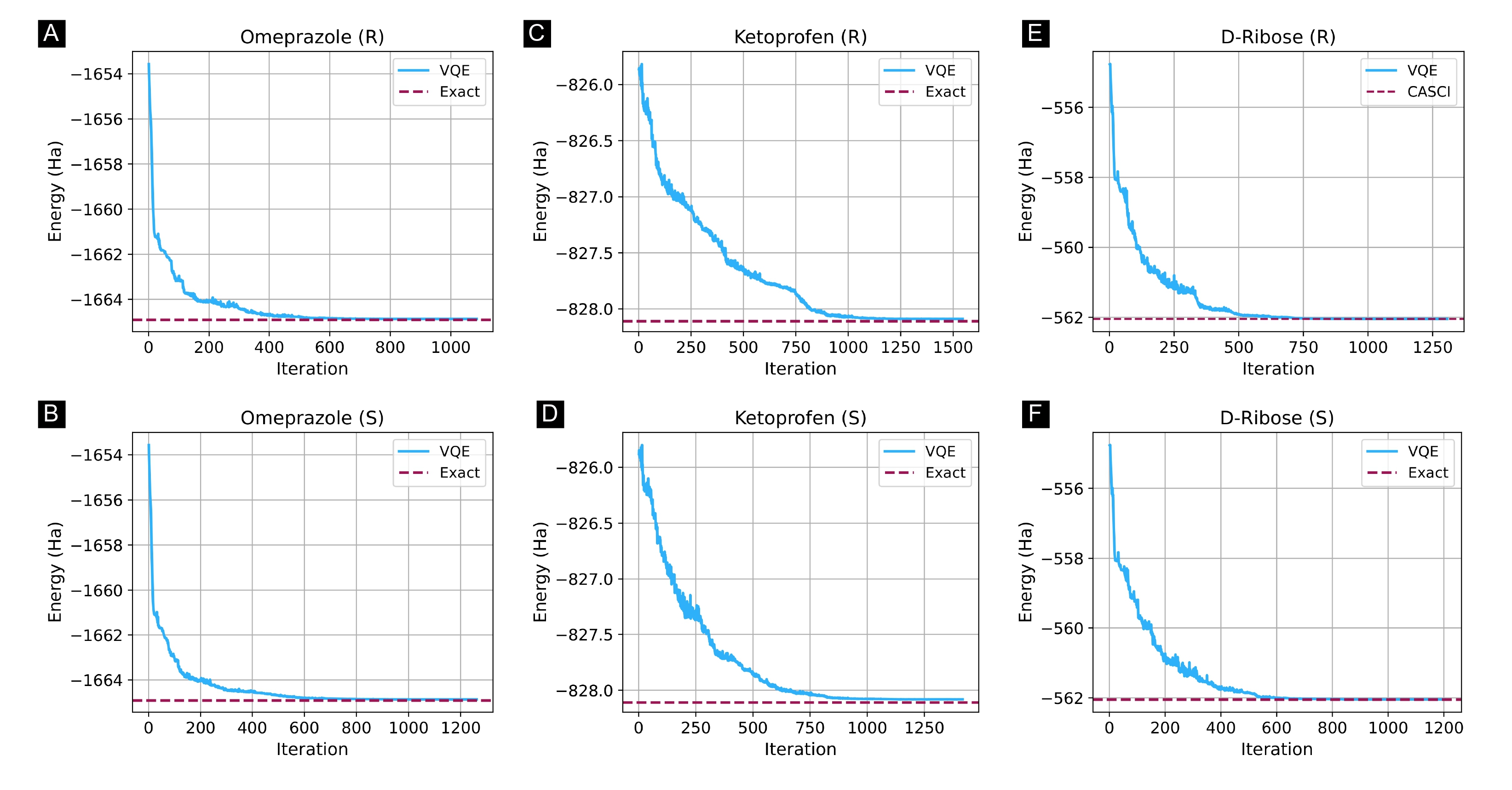}
\caption{\textbf{Convergence of the variational quantum eigensolver (VQE) ground-state energy relative to the exact reference solution for representative chiral molecules.}  
 The top row shows the ground-state energy (in Hartree) as a function of the number of optimization iterations for the (a) R-omeprazole, (b) R-ketoprofen, and (c) R-D-ribose enantiomers. The bottom row presents the corresponding convergence behavior for the S-enantiomers of each molecule. In all panels, the VQE energies are compared against exact solver results, illustrating the convergence characteristics and accuracy of the VQE approach across different chiral systems.}
\label{gse}
\end{figure*}

Let $\lvert \Psi_0 \rangle$ denote the VQE-optimized ground state and $\lvert \Psi_\mu \rangle$ an excited state obtained from the qEOM formalism with excitation energy $\omega_\mu$. The electric transition dipole moment between the ground and excited states is formally defined as
\begin{equation}
\boldsymbol{\mu}_{0\mu}
=
\langle \Psi_0 \lvert \hat{\boldsymbol{\mu}} \rvert \Psi_\mu \rangle ,
\end{equation}
where $\hat{\boldsymbol{\mu}} = -\sum_i \mathbf{r}_i$ is the electronic electric dipole operator. Analogously, the magnetic transition dipole moment is given by
\begin{equation}
\boldsymbol{m}_{0\mu}
=
\langle \Psi_0 \lvert \hat{\boldsymbol{m}} \rvert \Psi_\mu \rangle ,
\end{equation}
where $\hat{\boldsymbol{m}} = -\tfrac{1}{2} \sum_i \left( \mathbf{r}_i \times \mathbf{p}_i \right)$ denotes the orbital magnetic dipole operator. Here, $\mathbf{r}_i$ and $\mathbf{p}_i$ denote the position and momentum operators of the $i$-th electron.

Within the qEOM framework, the excited states are represented as linear combinations of excitation operators acting on the correlated ground state,
\begin{equation}
\lvert \Psi_\mu \rangle
=
\hat{O}_\mu^\dagger
\lvert \Psi_0 \rangle
=
\sum_{\nu}
c_{\nu}^{(\mu)}
\,
\hat{E}_{\nu}^{\dagger}
\lvert \Psi_0 \rangle ,
\end{equation}
where $\{ \hat{E}_{\nu}^{\dagger} \}$ denotes the truncated singles--doubles excitation operator basis and $c_{\nu}^{(\mu)}$ are the expansion coefficients obtained from solving the qEOM generalized eigenvalue problem.

In practice, transition moments are evaluated without explicit preparation of excited states by expressing them in terms of commutators with the qEOM excitation operators. For a given excited state $\mu$, the Cartesian component $\alpha \in \{x,y,z\}$ of the electric transition dipole moment is computed as
\begin{equation}
\mu_{0\mu}^{(\alpha)}
=
-
\frac{
\sum_{\nu}
c_{\nu}^{(\mu)}
\,
\left\langle \Psi_0 \right|
\left[
\hat{O}_{\mu},
\hat{E}_{\nu}^{\dagger}
\right]
\left| \Psi_0 \right\rangle
}{
\sqrt{
\left\langle \Psi_0 \right|
\left[
\hat{O}_{\mu},
\hat{O}_{\mu}^{\dagger}
\right]
\left| \Psi_0 \right\rangle
}
} ,
\end{equation}
where $\hat{D}^{(\alpha)}$ denotes the $\alpha$-component of the electric dipole operator expressed in the active-space basis,
\begin{equation}
\hat{D}^{(\alpha)}
=
\sum_{pq}
d_{pq}^{(\alpha)}
\,
\hat{a}_p^{\dagger} \hat{a}_q ,
\end{equation}
with $d_{pq}^{(\alpha)}$ being the one-electron electric dipole integrals.

Analogously, the magnetic transition dipole moment is evaluated as
\begin{equation}
m_{0\mu}^{(\alpha)}
=
\frac{
\sum_{\nu}
c_{\nu}^{(\mu)}
\,
\left\langle \Psi_0 \right|
\left[
\hat{M}^{(\alpha)},
\hat{E}_{\nu}^{\dagger}
\right]
\left| \Psi_0 \right\rangle
}{
\sqrt{
\left\langle \Psi_0 \right|
\left[
\hat{O}_{\mu},
\hat{O}_{\mu}^{\dagger}
\right]
\left| \Psi_0 \right\rangle
}
} ,
\end{equation}
where $\hat{M}^{(\alpha)}$ denotes the $\alpha$-component of the magnetic dipole operator in the active-space basis,
\begin{equation}
\hat{M}^{(\alpha)}
=
\sum_{pq}
m_{pq}^{(\alpha)}
\,
\hat{a}_p^{\dagger} \hat{a}_q ,
\end{equation}
and $m_{pq}^{(\alpha)}$ are the corresponding one-electron magnetic dipole integrals.

The rotatory strength associated with excitation $\mu$, which determines the intensity of electronic circular dichroism transitions, is finally evaluated as
\begin{equation}
R_{0\mu}
=
\mathrm{Im}
\left[
\boldsymbol{\mu}_{0\mu}
\cdot
\boldsymbol{m}_{0\mu}^{\ast}
\right] ,
\end{equation}
where $\mathrm{Im}[\cdot]$ denotes the imaginary part. The resulting excitation energies $\omega_\mu$ and rotatory strengths $R_{0\mu}$ fully characterize the electronic circular dichroism spectrum within the active-space approximation.

\section{Experimental and Computational Setup}
To validate the accuracy of the proposed framework, we employ  state-vector simulations that apply the exact unitary circuit evolution, thereby excluding both sampling errors and hardware-induced noise. The algorithm is evaluated on 12 chiral pharmaceutical molecules. The corresponding electronic Hamiltonians are derived ab intio using  one and two-electron integrals calculated at the Hartree–Fock level with the STO-3G basis.

\subsection{Study population}

The study population comprises fifteen chemically and therapeutically diverse chiral drug molecules selected to assess the generality and robustness of the proposed quantum-enabled framework for chiroptical property calculations. The molecular set includes nonsteroidal anti-inflammatory drugs, cardiovascular agents, central nervous system drugs, bronchodilators, anticoagulants, and immunomodulatory compounds, thereby spanning a broad range of molecular sizes, functional groups, and stereochemical complexity.

For each molecule, stereochemical information was obtained using RDKit-generated molecular representations. In the case of racemic drugs, achiral SMILES strings were employed, reflecting the absence of a single clinically dominant enantiomer. For enantiopure drugs, the clinically relevant stereoisomer was explicitly selected and encoded using chiral SMILES notation. This selection ensures that the investigated structures accurately represent experimentally relevant configurations and associated chiroptical responses.

Table~\ref{chiral_list} summarizes all single and multi-chiral-center molecules studied in this work, including their SMILES representations and associated therapeutic applications. This enables a systematic evaluation of the proposed approach across structurally diverse chiral pharmaceuticals and provides a representative benchmark for quantum-based electronic circular dichroism computations.

\begin{figure*}[!ht]
	\centering
\includegraphics[scale=0.3]{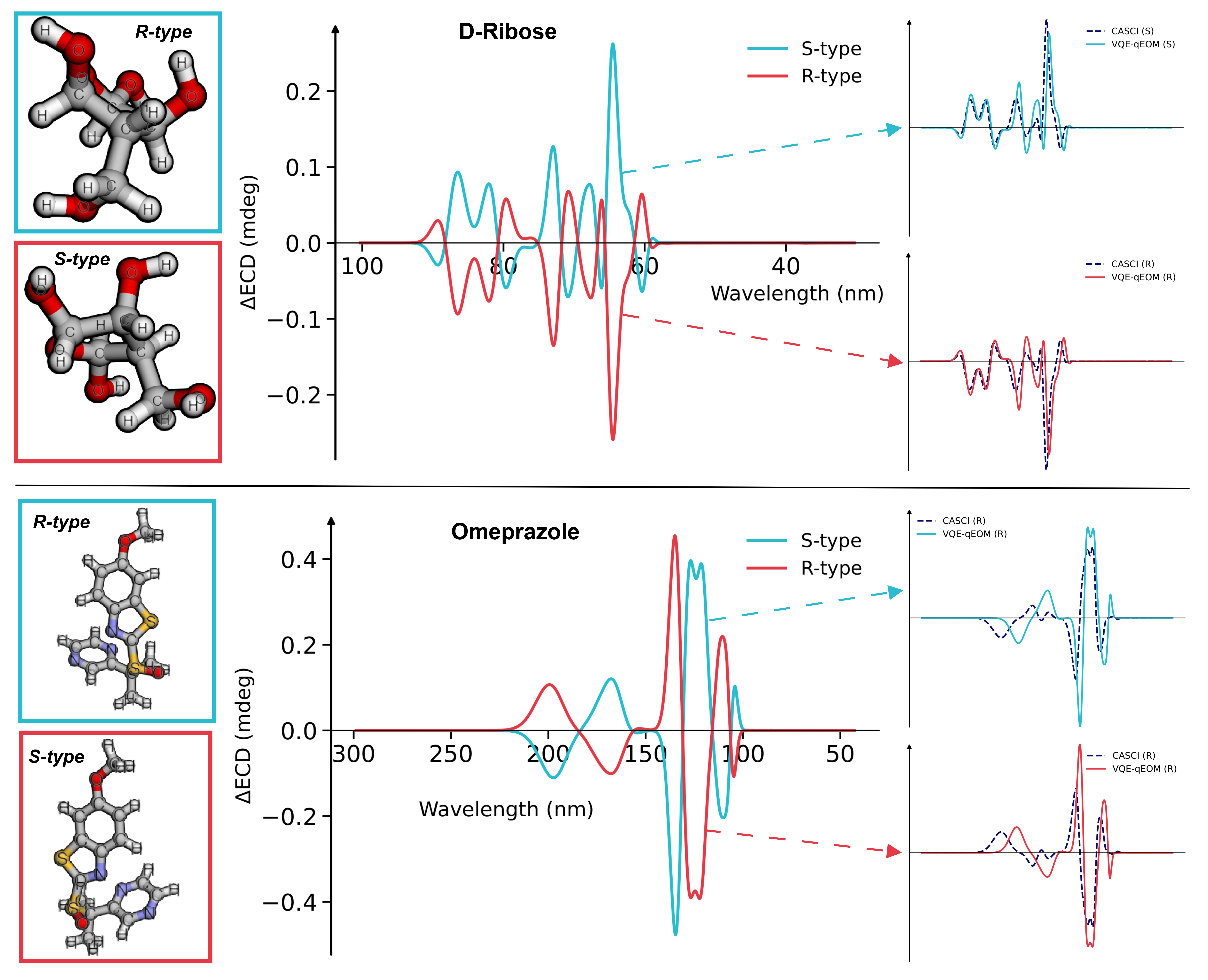}
\caption{\textbf{Quantum prediction of electronic circular dichroism in chiral molecules.}  Top: R- and S-enantiomers of D-ribose with ECD spectra computed using the VQE–qEOM framework, showing mirror-image responses and quantitative agreement with classical CASCI results in the CAS(12,12) active space mapped to 24 qubits. Bottom: ECD spectra of chiral omeprazole obtained with VQE–qEOM, exhibiting close agreement with CASCI reference calculations. The quantum results show very close agreement with classical CASCI reference calculations, accurately reproducing the sign, relative intensity, and spectral features of the ECD response.}
\label{mainfig28}
\end{figure*}

\subsection{Quantum circuit and simulation details}

In the initial classical pre-processing step, calculations are performed using PySCF and
include active-space selection based on natural orbital occupation numbers (NOON)
derived from MP2/cc-pVDZ calculations.
The resulting one- and two-electron Hamiltonian integrals are computed at the restricted
Hartree–Fock level and supplied to the VQE algorithm. Variational parameters are
initialized with small random values to seed the optimization. The fermionic Hamiltonian and wavefunction ansatz are encoded on qubit registers using
the Jordan-Wigner transformation. Under this mapping, the second-quantized fermionic
operators are systematically translated into sums of Pauli operator strings acting on
qubits, yielding a qubit representable Hamiltonian. The same transformation is applied
consistently to the unitary coupled-cluster ansatz, ensuring that both the Hamiltonian and
the parametrized wavefunction are expressed within an identical qubit operator framework.

\begin{figure*}[!ht]
	\centering
\includegraphics[scale=0.44]{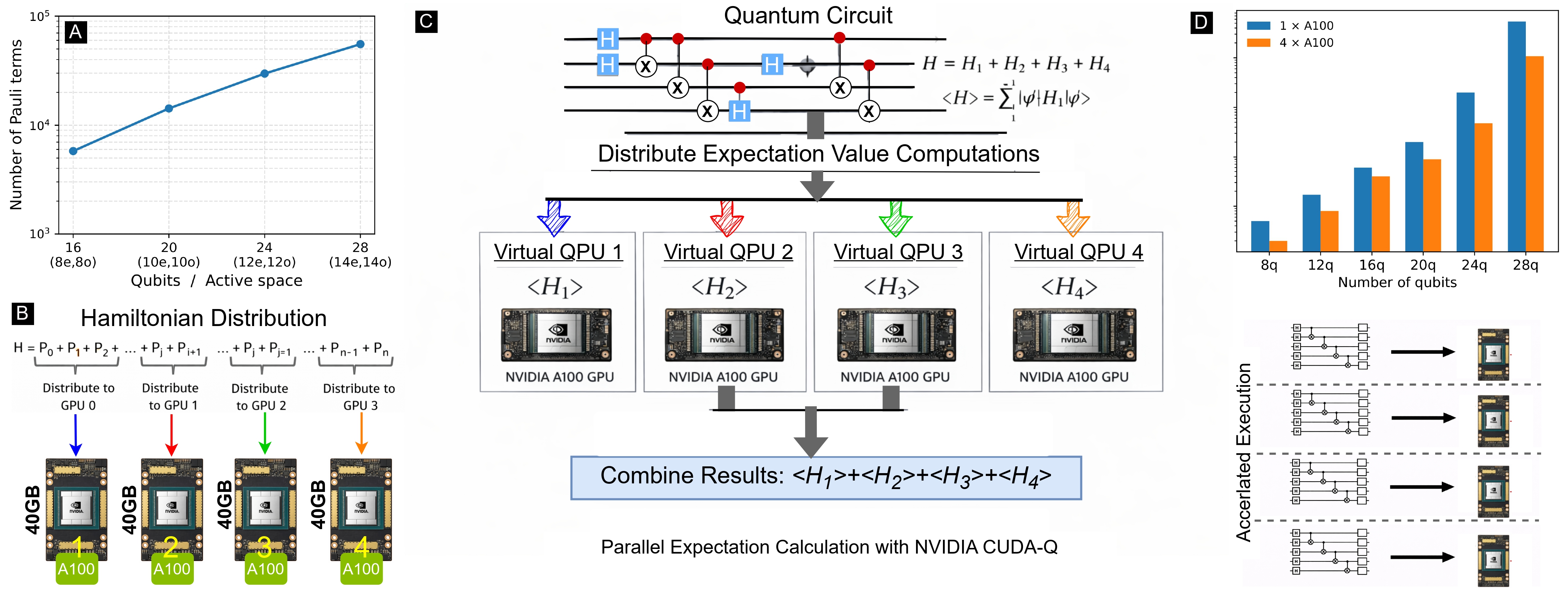}
\caption{\textbf{Scaling and parallelization of quantum Hamiltonian expectation evaluation on multi-GPU platforms.}  Top: (A) Scaling of the number of Pauli terms with increasing qubit count, highlighting the computational bottleneck in large active-space simulations. (B) Partitioning of the Hamiltonian into subsets of Pauli terms distributed across four NVIDIA A100 GPUs (40 GB each). (C) Parallel evaluation of expectation values on four virtual QPUs, with partial results combined to obtain the total Hamiltonian expectation. (D) Bar-plot comparison of runtime, demonstrating the speedup achieved using four virtual QPUs relative to a single QPU.}
\label{gpu}
\end{figure*}

The variational wavefunction is prepared by initializing the quantum register in a
Hartree--Fock reference state, constructed by applying Pauli-$X$ gates to qubits
corresponding to occupied active-space orbitals. Correlation effects are introduced through a parametrized quantum circuit consisting of
three repeated ansatz layers, each comprising single-qubit $R_y$ rotations followed by a
fully entangling network of controlled-NOT gates acting across the qubit register. This
layered construction produces a highly entangled trial wavefunction whose parameters are
optimized variationally. Following state preparation, the energy is evaluated by measuring the expectation value
of the qubit-mapped Hamiltonian ($\hat{H}_{\mathrm{act}}$). The simulations are performed using NVIDIA’s multi-GPU statevector simulator implemented within the CUDA-Q framework. A 24-qubit register is used for the CAS(12,12) active space, corresponding to 12 electrons in 12 spatial orbitals, while a 20-qubit register is employed for the CAS(10,10) active space, corresponding to 10 electrons in 10 spatial orbitals. For the CAS(10,10) active space, the excited-state manifold is constructed using the complete set of single and double excitation operators. In contrast, for the larger CAS(12,12) active space, the excited-state space is restricted to the lowest 100 excited configurations, selected to manage the computational cost while retaining the dominant contributions to the low-lying excited states. As shown in Fig.~\ref{gse}, the VQE ground-state energies converge rapidly toward the exact reference solutions for all chiral molecules considered. Accurate energies are obtained using only two to three layers of the variational ansatz for both R and S-enantiomers. This demonstrates the capability of shallow VQE circuits to efficiently capture molecular ground-state properties.

To address the rapid growth of Pauli terms with increasing qubit count, we distribute the Hamiltonian across four NVIDIA A100 GPUs and evaluate expectation values in parallel using four virtual QPUs, as shown in Fig~\ref{gpu}. Each virtual QPU computes a disjoint subset of Pauli terms, and the partial expectations are aggregated to obtain the full Hamiltonian expectation value. This strategy yields a substantial runtime reduction compared with a single-QPU execution, enabling scalable simulations in large active spaces.

In practice, the excitation operators are generated in a fully spin-resolved
spin-orbital basis. For a closed-shell reference, each spatial orbital gives rise
to $\alpha$ and $\beta$ spin orbitals, and all allowed spin combinations consistent
with fermionic antisymmetry are included. As a result, the excitation manifold
contains both same-spin and mixed-spin contributions, which is essential for a
balanced description of singlet and triplet excited states.

For an $(N_e,N_o)$ active space, the number of occupied and virtual spatial orbitals
are $n_{\mathrm{occ}}=N_e/2$ and $n_{\mathrm{vir}}=N_o-N_e/2$, respectively. The total
dimension of the qEOM excitation space scales combinatorially with the active-space
size and is given by
\begin{equation}
N_{\mathrm{exc}}
=
2\,n_{\mathrm{occ}}\,n_{\mathrm{vir}}
\;+\;
2\binom{n_{\mathrm{occ}}}{2}\binom{n_{\mathrm{vir}}}{2}
\;+\;
n_{\mathrm{occ}}^2 n_{\mathrm{vir}}^2,
\end{equation}
where the first term corresponds to spin-resolved single excitations, the second
term to same-spin double excitations, and the third term to mixed-spin doubles.
This excitation basis is subsequently mapped to qubit operators via a
Jordan--Wigner transformation and used to assemble the qEOM matrices through
nested commutators with the electronic Hamiltonian.

As a concrete example, an $(8e,8o)$ active space yields $n_{\mathrm{occ}}=4$ and
$n_{\mathrm{vir}}=4$, resulting in 32 single and 328 double excitations, for a total
of 360 qEOM operators.

The variational parameters $\boldsymbol{\theta}$ are optimized using the classical
COBYLA algorithm. The optimization is terminated once the total energy satisfies a
predefined convergence threshold, at which point the converged energy is recorded. For the CAS(12,12) active space, the VQE calculations are performed using a 24-qubit statevector representation. The Hamiltonian is distributed across four NVIDIA A100 GPUs within the CUDA-Q framework, enabling parallel evaluation of Hamiltonian expectation values during the VQE optimization.

With access to the converged variational ground state $\lvert \Psi_0 \rangle$ obtained from
statevector simulations, excited-state energies are computed using the qEOM formualism. Excited states are generated by acting on
$\lvert \Psi_0 \rangle$ with a set of excitation operators $\hat{O}_u^\dagger$ comprising both
single and double excitations within the active space, as defined in Eqs.~(15--18). Solving the resulting qEOM
eigenvalue problem yields the electronic excitation energies relative to the ground state.

On the basis of these excited states, the electric and magnetic transition dipole moments, $\boldsymbol{\mu}_{0\mu}$ and $\boldsymbol{m}_{0\mu}$, are evaluated within the qEOM framework using ground-state expectation values, without explicit preparation of excited-state wavefunctions, as defined in Eqs.~(28--31). These transition moments are combined to compute the corresponding rotatory strengths
$R_{0\mu}$, as given in Eq.~(32), which, together with the excitation energies $\omega_\mu$,
fully determine the electronic circular dichroism spectra of the chiral drug molecules
considered in this study.

The results are benchmarked against exact diagonalization, CASCI, and CCSD calculations performed within the same active orbital space. To ensure consistency across all molecular systems studied, all VQE simulations employ an identical variational ansatz. A uniform convergence threshold of $\epsilon = 10^{-4}$ is used throughout, together with the minimal STO-3G basis set.
The use of a minimal basis is motivated by current noisy intermediate-scale quantum
(NISQ) hardware constraints, as larger basis sets would significantly increase both the
qubit count and circuit depth.


\begin{figure*}[!ht]
	\centering
\includegraphics[scale=0.75]{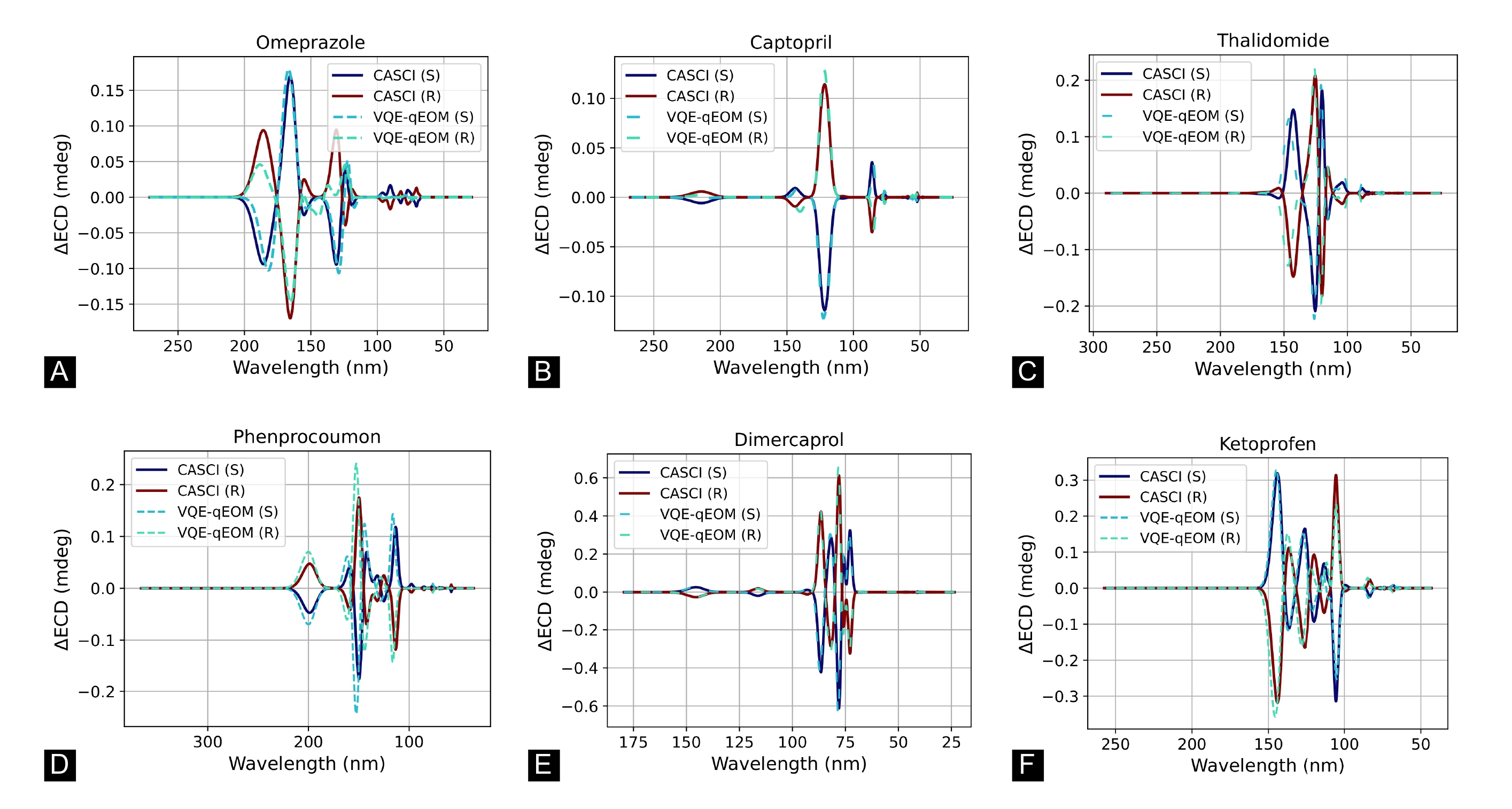}
\caption{\textbf{Electronic circular dichroism spectra of single–chiral-center drug molecules computed using CASCI and VQE–qEOM methods.}  Panels (a–f) show representative molecules (Omeprazole, Captopril, Thalidomide, Phenprocoumon, Dimercaprol, and Ketoprofen), where solid lines correspond to CASCI results and dashed lines to VQE–qEOM predictions. For each molecule, the ECD spectra of the $R$ and $S$ enantiomers display the expected mirror symmetry, with VQE–qEOM accurately reproducing both the enantiomeric sign inversion and the overall spectral features observed in the classical reference calculations. These results demonstrate that VQE–qEOM reliably captures chiroptical response in pharmaceutically relevant single-stereocenter systems.}
\label{single16}
\end{figure*}

\subsection{Classical reference calculations}
Accurately describing correlated electronic states in chiral molecular systems requires
computational approaches that balance chemical accuracy with tractable computational cost.
This balance remains difficult to achieve with classical simulations, as exact methods such as
full configuration interaction (FCI) exhibit factorial scaling with system size. Even when the
correlation treatment is restricted to an active space, CASCI scales combinatorially with the number of active electrons and orbitals,
leading to an exponential growth of the configuration space. Widely used approximate
wavefunction approaches, including coupled-cluster singles and doubles (CCSD) and its
perturbative triples extension CCSD(T), scale as $\mathcal{O}(N^6)$ and $\mathcal{O}(N^7)$,
respectively, where $N$ denotes the number of molecular orbitals. Consequently, the
application of these methods to chemically realistic chiral drug molecules, particularly when
extended active spaces are required, rapidly becomes computationally prohibitive.

To validate the proposed VQE–qEOM framework for ECD prediction, classical reference calculations are performed using the PySCF package with the STO-3G basis set. The same active spaces employed in the quantum simulations are used consistently across all classical benchmarks.

Excited-state energies and ECD spectra are evaluated using CASCI, which corresponds to a full configuration-interaction treatment within the chosen active space and therefore provides an exact reference for the selected orbitals. Coupled-cluster calculations at the CCSD level are used exclusively to obtain accurate ground-state reference energies, with single and double excitations restricted to the same active space. Excited-state and chiroptical properties are not computed at the CCSD level. Consequently, ground-state energies are benchmarked against CCSD and CASCI, whereas excited-state energies and ECD responses are validated against CASCI. As shown in Table~\ref{gse}, ground-state energies predicted by VQE–qEOM closely match CASCI and CCSD reference values for all chiral drug molecules examined, validating the accuracy of the quantum framework at the ground-state level.

\begin{figure*}[!ht]
	\centering
\includegraphics[scale=0.75]{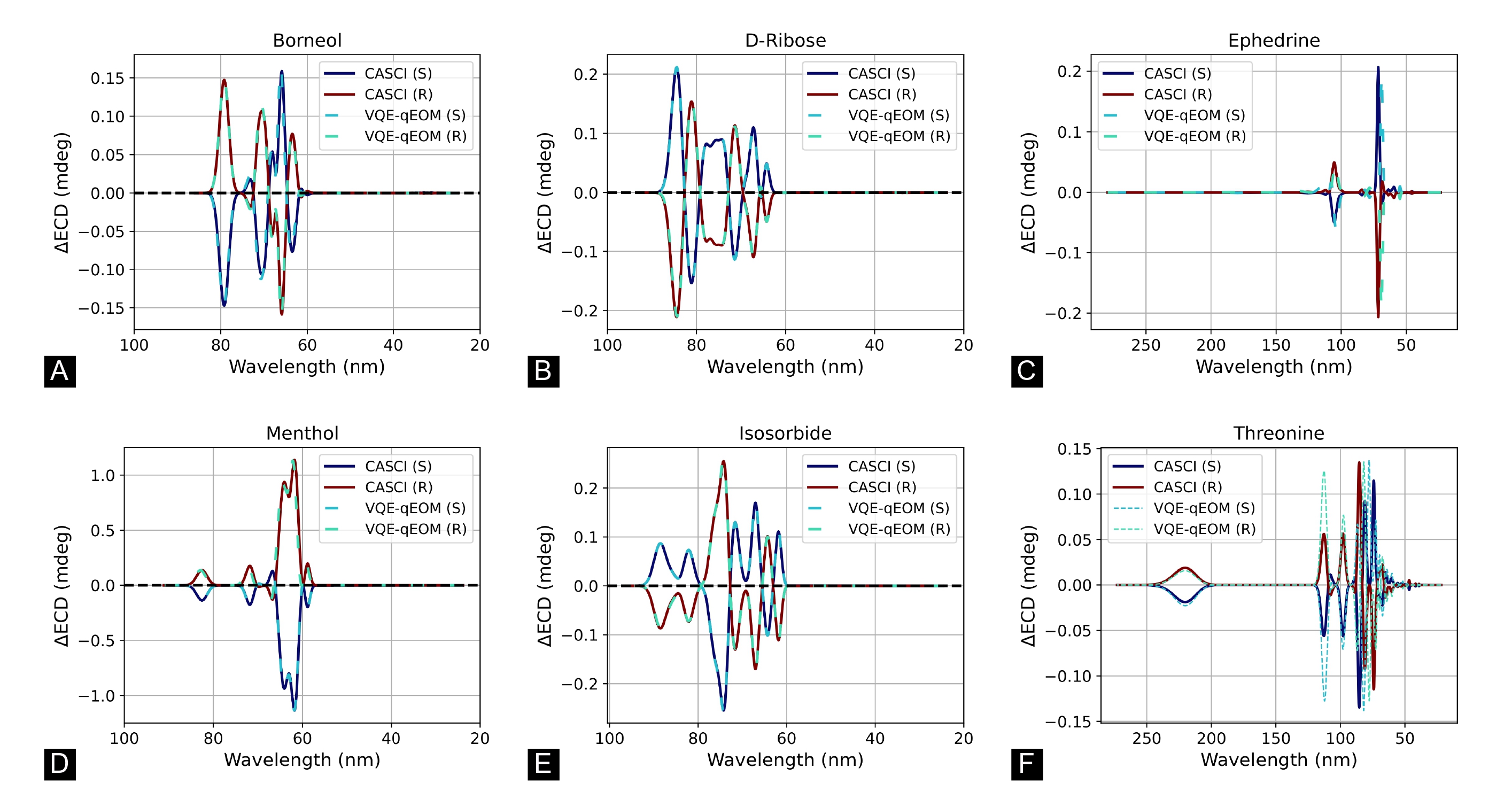}
\caption{\textbf{Electronic circular dichroism (ECD) spectra of multi–chiral-center drug molecules computed using CASCI and VQE–qEOM methods.} Panels (a–f) show representative molecules with multiple stereogenic centers (Borneol, D-Ribose, Ephedrine, Menthol, Isosorbide, and Threonine), where solid lines correspond to CASCI results and dashed lines to VQE–qEOM predictions. For each molecule, the ECD spectra of the $R$/$S$ enantiomeric configurations exhibit the expected chiroptical behavior, with VQE–qEOM closely matching the classical CASCI reference across the full spectral range. Near-quantitative agreement is observed for all systems, with only minor deviations in peak intensity for threonine, while preserving the correct spectral shape and sign. These results further demonstrate the robustness of VQE–qEOM in accurately capturing chiroptical responses of pharmaceutically relevant molecules with multiple chiral centers.}
\label{multi16}
\end{figure*}

\section{Results and Discussion}
We evaluate the accuracy of the VQE–qEOM framework by comparing calculated ECD spectra against classical reference data, focusing on the positions of spectral bands, the sign of the Cotton effects, their relative intensities, and the overall spectral agreement. Fig.~\ref{mainfig28} demonstrates the capability of the proposed VQE–qEOM framework to deliver accurate ECD predictions in large active spaces while maintaining computational efficiency. For the prototypical chiral molecule D-ribose, the quantum results obtained in the CAS(12,12) active space (24 qubits) show quantitative agreement with classical CASCI, validating the accuracy of the excited-state quantum workflow. Importantly, the same framework achieves consistent and reliable ECD predictions for the structurally more complex pharmaceutical omeprazole, highlighting its robustness, accelerated runtime, and scalability for realistic chiral molecules.

Next, we assess the performance of the 20-qubit VQE–qEOM framework for predicting chiroptical responses in molecules containing a single stereogenic center. Fig~\ref{single16} presents ECD spectra for a set of pharmaceutically relevant compounds, comparing variational quantum predictions with classical active-space references. Across all systems examined, the quantum-derived spectra reproduce the characteristic mirror symmetry between enantiomers as well as the dominant spectral features of the classical calculations, demonstrating that the VQE–qEOM approach captures the essential excited-state physics governing ECD in single-chiral-center molecules. 

The demonstrated accuracy of the VQE–qEOM framework for single-chiral-center molecules motivates its extension to systems with multiple chiral centers, where coupled stereochemical effects and increased electronic complexity provide a more stringent test. Fig~\ref{multi16} examines the performance of the VQE–qEOM framework for molecules containing multiple stereogenic centers, a regime characterized by increased electronic complexity and coupled chiral effects. Across the systems considered, the quantum-derived ECD spectra closely reproduce the classical active-space references over the full spectral range, capturing both the sign and relative intensity of the dominant Cotton effects. Minor deviations in peak intensity are observed for threonine, but the overall spectral shape and enantiomeric behavior remain consistent with the classical results. These findings indicate that the VQE–qEOM approach remains robust and quantitatively reliable for pharmaceutically relevant molecules with multiple chiral centers. The minor intensity deviations observed for threonine likely arise from enhanced sensitivity to correlation and conformational effects in more flexible molecules, rather than from a breakdown of the underlying variational quantum description. Overall, these results indicate that the accuracy of the VQE–qEOM approach is controlled primarily by the quality of the active-space representation, and remains stable as stereochemical complexity increases.

\begin{figure*}[!ht]
	\centering
\includegraphics[scale=0.7]{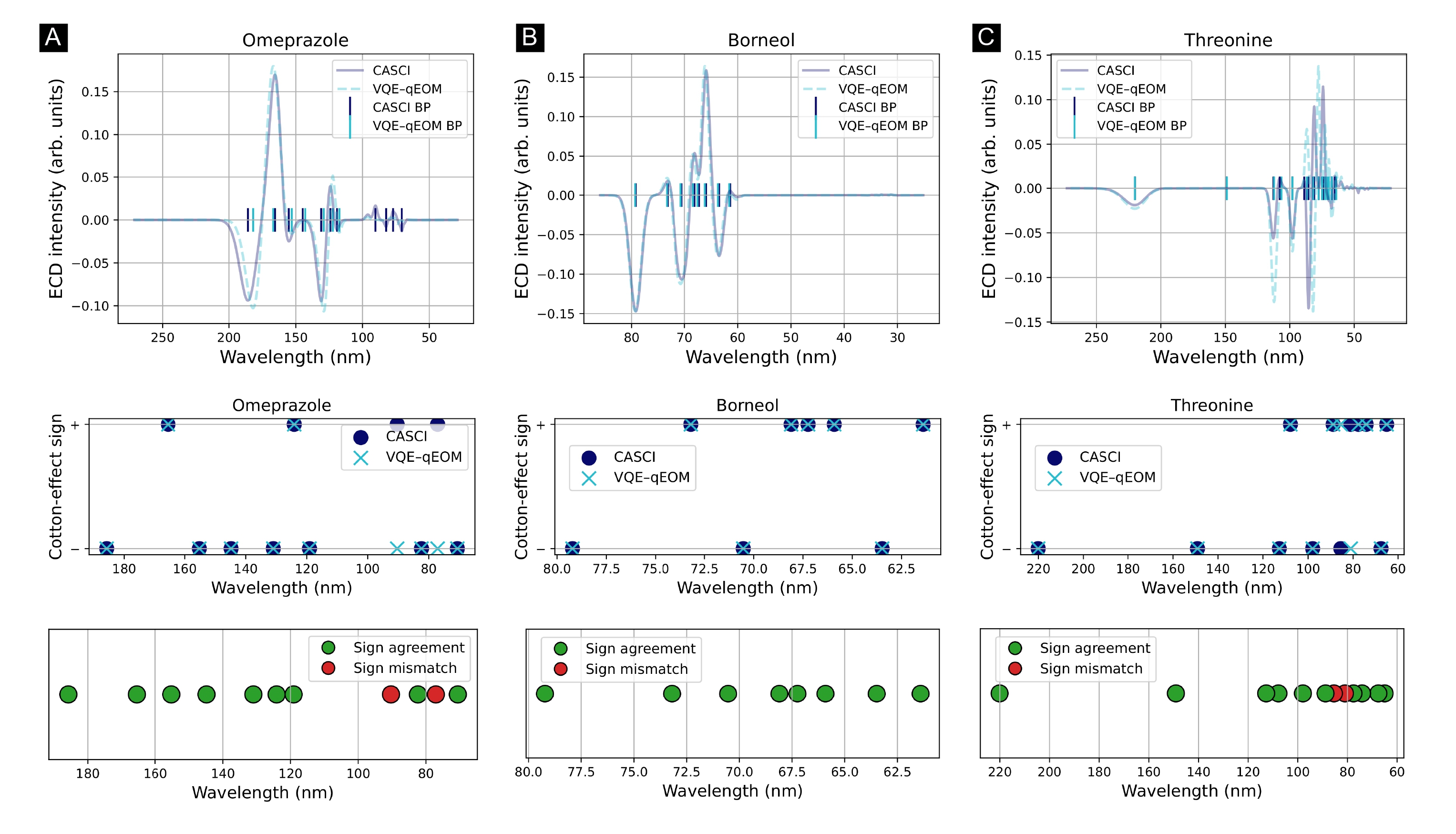}
\caption{
\textbf{Comparison of ECD band positions and Cotton-effect signs between CASCI and VQE--qEOM for representative chiral molecules.}
Panels (A)--(C) show results for Omeprazole, Borneol, and Threonine, respectively.
For each molecule, the positions of the dominant ECD bands, the sign of the corresponding Cotton effects, and a strip plot summarizing sign agreement are shown.
Borneol exhibits perfect agreement with the CASCI reference, with all band positions and Cotton-effect signs accurately reproduced by VQE--qEOM.
Omeprazole shows overall excellent agreement, with only minor deviations in band alignment.
For Threonine, VQE--qEOM reproduces the majority of Cotton-effect signs, with a small number of mismatches observed in the short-wavelength region (70--80~nm), while preserving the overall spectral trends.
}
\label{cottoneffect}
\end{figure*}

To further examine the fidelity of the VQE–qEOM predictions beyond qualitative spectral agreement, we analyze the alignment of ECD band positions and the associated Cotton-effect signs relative to classical active-space references. Fig~\ref{cottoneffect} provides a quantitative comparison of ECD band positions and Cotton-effect signs obtained from VQE–qEOM and classical CASCI calculations for representative chiral molecules. Across all systems examined, the variational quantum approach correctly reproduces the sign of the dominant Cotton effects associated with the lowest lying electronic excitations, which are the primary contributors to the overall ECD response. The near-perfect agreement observed for borneol reflects its relatively rigid molecular framework and well-separated excited-state manifold, which is readily captured within the chosen active space. In contrast, the small discrepancies in band alignment for omeprazole and the limited sign mismatches observed for threonine at short wavelengths are consistent with increased sensitivity to higher-energy excitations and correlation effects that lie closer to the boundary of the active-space description. Importantly, these deviations do not alter the overall spectral trends or enantiomeric assignments, underscoring the robustness of the VQE–qEOM framework for chiroptical analysis.

\begin{figure*}[!ht]
	\centering
\includegraphics[scale=0.91]{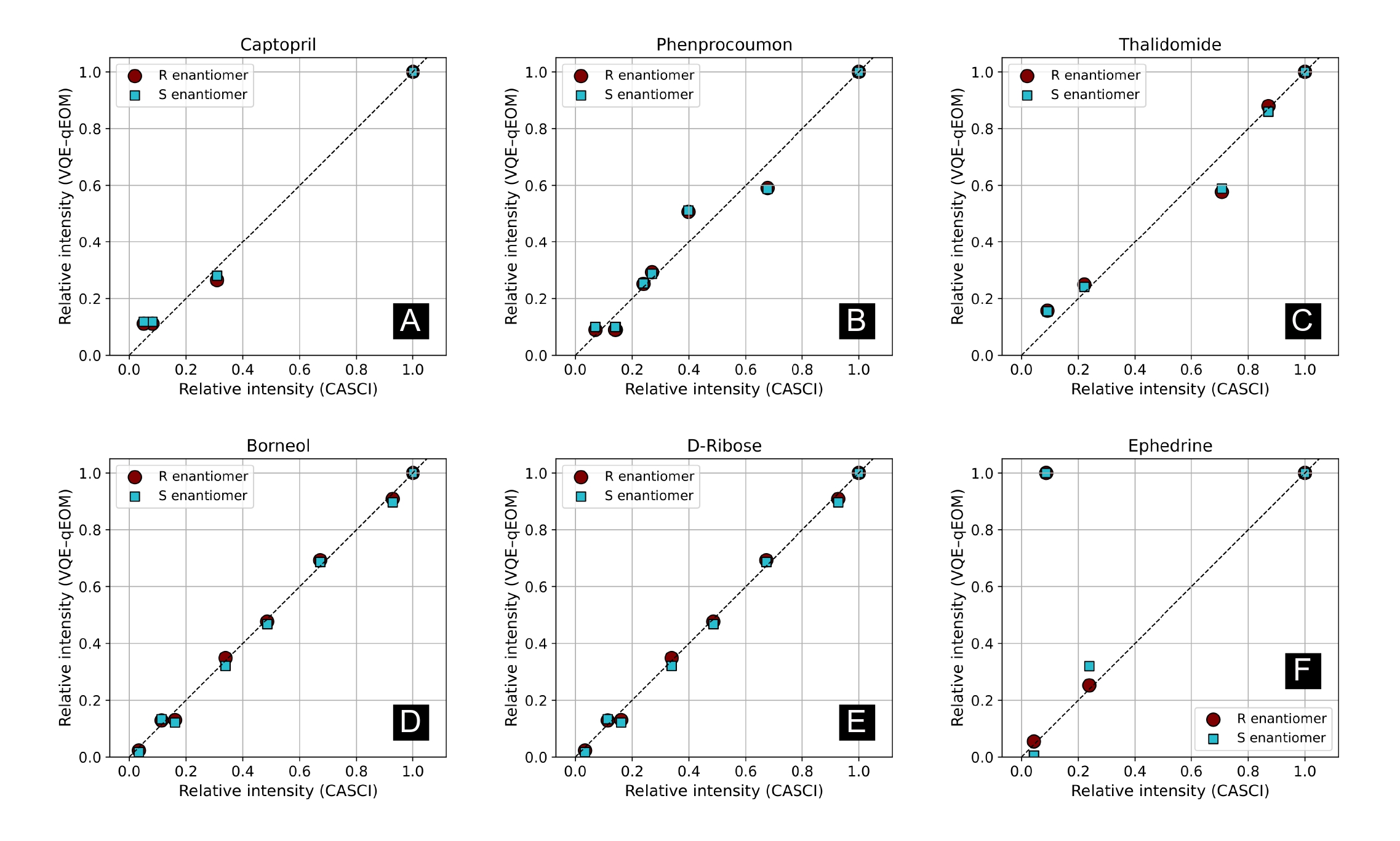}
\caption{
\textbf{Relative ECD band intensity comparison between CASCI and VQE--qEOM for single- and multi-chiral-center molecules.}
Panels (A)--(C) show single–chiral-center systems (Captopril, Phenprocoumon, and Thalidomide), while panels (D)--(F) correspond to multi–chiral-center molecules (Borneol, D-ribose, and Ephedrine).
Each marker represents a dominant ECD band, with intensities normalized to the strongest transition for each molecule; the dashed diagonal line indicates perfect agreement between CASCI and VQE--qEOM.
Captopril, Borneol, and D-ribose exhibit near-perfect alignment along the diagonal, demonstrating excellent agreement in relative band intensities.
For Phenprocoumon and Thalidomide, the majority of bands closely follow the reference line, with only minor deviations for a small number of transitions, while preserving the overall intensity trends observed in CASCI.
In contrast, Ephedrine shows larger deviations for several bands, consistent with its very narrow ECD spectral features, which amplify sensitivity to small shifts in excitation energies and lead to noticeable differences in relative peak intensities between the classical and quantum spectra.
}

\label{relativeins}
\end{figure*}

Beyond reproducing spectral shapes and Cotton-effect signs, quantitative agreement in relative ECD band intensities provides a stringent test of excited-state accuracy. Fig~\ref{relativeins} compares normalized band intensities obtained from VQE–qEOM with classical CASCI references for both single- and multi-chiral-center molecules. For most systems, the quantum-derived intensities closely follow the CASCI benchmarks, indicating that the variational quantum framework accurately captures the relative transition strengths governing the dominant chiroptical response. The near-perfect agreement observed for captopril, borneol, and D-ribose reflects well-separated low-lying excitations whose transition characters are robust within the chosen active spaces. In contrast, the larger deviations seen for ephedrine are consistent with its narrow and closely spaced ECD features, where small differences in excitation energies lead to amplified changes in normalized peak intensities. Importantly, even in this more sensitive regime, the overall intensity ordering and spectral trends remain preserved, underscoring the robustness of the VQE–qEOM approach across chemically diverse chiral systems.

Overall, our results demonstrate that the VQE–qEOM framework quantitatively reproduces classical active-space references across a diverse set of chiral drug molecules, capturing ground-state energetics, excited-state energies, ECD band positions, Cotton-effect signs, and relative band intensities for both single- and multi-chiral-center systems. The method preserves the expected enantiomeric mirror symmetry and accurately describes the dominant low-lying excitations that govern chiroptical response, with only minor deviations observed in systems exhibiting dense or narrowly spaced spectral features. Together, these findings establish VQE–qEOM as a robust and scalable approach for first-principles ECD prediction and chiral assignment, providing a practical foundation for extending quantum simulations of chiroptical properties to larger molecules and more complex stereochemical environments as quantum hardware continues to advance.

\section{Conclusion and Outlook}
In this work, we present an efficient variational quantum framework for predicting electronic circular dichroism (ECD) spectra of chiral drug molecules, enabling direct characterization of left- and right-handed enantiomers through their excited-state and chiroptical properties. By integrating the variational quantum eigensolver with the quantum equation-of-motion formalism within a multi-GPU- and QPU-accelerated hybrid quantum–classical workflow, electronic excitation energies and transition properties are obtained from a correlated ground-state reference. The approach is demonstrated using chemically relevant active spaces mapped to quantum circuits comprising approximately 20–24 qubits, extending beyond minimal models while remaining compatible with near-term quantum hardware. This qEOM-based treatment provides a fully quantum-consistent route to ECD simulation at increased active-space resolution.

The framework combines the robustness of variational ground-state methods with the conceptual structure of equation-of-motion techniques in a fully quantum formulation, while benefiting from efficient expectation-value evaluation on quantum simulators. Across all chiral drug molecules investigated, the predicted ECD spectra closely match CASCI benchmarks, accurately reproducing characteristic chiroptical signatures. Collectively, these results indicate that the framework reliably predicts ECD spectra for both 
R and 
S-enantiomeric pairs, particularly in cases where the spectra exhibit pronounced mirror symmetry.

The strong agreement observed for molecules with multiple chiral centers reflects the fact that the dominant ECD response is governed by a limited number of low-lying excited states, which are consistently captured within the selected active-space representation. By treating these states on an equal footing with classical CASCI references, the VQE–qEOM framework preserves excitation ordering and transition character, thereby determining Cotton-effect sign and magnitude. These results establish a practical pathway for scaling quantum simulations of chiroptical properties to larger and more complex molecular systems, with direct relevance to stereochemical analysis and chiral drug discovery.

\section*{Acknowledgments}

The authors would like to acknowledge the financial support from the Quantum Science Center, a National Quantum Information Science Research Center of the U.S. Department of Energy (DOE), operated at Oak Ridge National Laboratory
(ORNL)
	
	\bibliography{mainfile}
	
\end{document}